\documentclass[]{aastex62}

\begin{document}
\title{\large{Dynamical Masses of Young Stars Inferred from Two Transitions of CO with ALMA}}
\author{Pranav H. Premnath}
\affiliation{Department of Astronomy, The University of Texas at Austin, Austin, TX 78712, USA}
\author{Ya-Lin Wu}
\affiliation{Department of Physics, National Taiwan Normal University, Taipei City, Taiwan}
\author{Brendan P. Bowler}
\affiliation{Department of Astronomy, The University of Texas at Austin, Austin, TX 78712, USA}
\author{Patrick D. Sheehan}
\affiliation{Department of Physics and Astronomy, Northwestern University, Evanston, IL 60208, USA}

\begin{abstract}
Stellar masses are fundamental but often difficult to measure. Thanks to the Atacama Large Millimeter Array (ALMA) and \textit{Gaia}, dynamical masses of pre-main sequence stars can be precisely measured using the Keplerian rotation of protoplanetary disks. We used ALMA CO(2-1) and CO(3-2) observations of CT Cha and DS Tau to determine their masses by modeling the geometry, kinematics, and physical properties of their disks with a Bayesian-based radiative transfer modeling code (\texttt{pdspy}). We found that the posterior distributions of the masses from the two transitions are inconsistent at the 2-4 $\sigma$ level. These systematic errors may originate from assumptions in the disk model, or perhaps the modest spatial or spectral resolutions used in this study. Regardless, this indicates that dynamical mass measurements using disk kinematics should be treated with caution when using only a single transition line because of these systematic errors.      
\end{abstract}

\section{Introduction}
The mass of a star is the most fundamental characteristic that governs its life cycle. Measuring stellar masses helps trace pathways in the H-R diagram and test theoretical stellar models. These evolutionary models are more uncertain at young ages as a result of changes in stellar rotation, radii, and magnetic field strength. Keplerian rotation in protoplanetary disks provides a means to directly measure stellar masses for pre-main sequence stars and test evolutionary models. Before \textit{Gaia} launched, disk-based dynamical masses often had uncertainties of 20\% or higher (e.g., \citealt{simon2000}; \citealt{hillenbrand2004}), which were typically dominated by uncertainties in distances to the stars. In recent years, the revolutionary power of ALMA and \textit{Gaia} has enabled precise dynamical mass measurements as small as a few percent (e.g., \citealt{czekala2015}; \citealt{simon2019}; \citealt{sheehan2019}). Of particular importance has been ALMA's unprecedented sensitivity to molecular line emission, which was limited only to the brightest sources for previous generations of millimeter interferometers.

The goal of this study is to compare the disk-based mass measurements from two CO transitions to assess their mutual consistency. We chose two pre-main sequence stars with bright, well-studied disks, CT Cha and DS Tau, for our study. Both stars are young (2-3 Myr; \citealt{simon2017}; \citealt{sheehan2019}) and nearby (191.8 $\pm$ 0.8 pc and 159.1 $\pm$ 1.1 pc, respectively; \citealt{gaia2018}).

The dynamical masses of both stars have been measured before with ALMA data. \citet{sheehan2019} computed the central mass of CT Cha to be $0.796^{+0.015}_{-0.014}~M_\odot$ using the CO(2-1) transition, and \citet{lodato2019} reported a mass of $0.83  \pm 0.02~M_\odot$ for DS Tau, using CO(3-2). Here we re-derive the masses of both stars using both CO(2-1) and CO(3-2) to test how modeling of different transitions can affect our mass measurements. We also compare our mass estimates to literature results to see how different models can affect the results. 

\newpage
\section{Analysis and Results}
We used ALMA Band 7 data (2016.1.01018.S., PI: Bowler) of CO(3-2) line emission at 345.796 GHz and ALMA Band 6 data (2015.1.00773.S., PI: Wu) of CO(2-1) line emission at 230.538 GHz to derive the dynamical mass of CT Cha. The velocity resolution is 0.85 km s$^{-1}$ for CO(3-2) and 0.32 km s$^{-1}$ for CO(2-1). Details of the observations can be found in \citet{wu2017} and \citet{wu2020}. After standard data reduction with CASA (\citealt{mcmullin2007}), we used \texttt{pdspy} (\citealt{sheehan2019}) to model the gas kinematics. Model fits were run using the supercomputer Lonestar 5 at the Texas Advanced Computing Center independently for the two transitions. \texttt{pdspy} uses an MCMC method to sample posterior probability distributions of 13 parameters in our disk model (see \citealt{sheehan2019} for details).

The CO(2-1) line yields a mass of $0.85 \pm 0.02~M_\odot$. This is $0.05 \pm 0.03~ M_\odot$ higher than that derived by \citet{sheehan2019}. On the other hand, the CO(3-2) line yields a mass of $0.96 \pm 0.02~  M_\odot$. Moreover, our inferred masses differ by $0.11 \pm 0.03 ~M_\odot$, or at the 3.9 $\sigma$ level. All values are reported after clipping out the potential outliers 4 $\sigma$ from the mean value. Note that the uncertainties represent one standard deviation from the mean of the posterior mass distribution. The left panel of Figure 1 displays the mass distributions of CT Cha.

\begin{figure}%
    \centering
    {\includegraphics[scale=0.5]{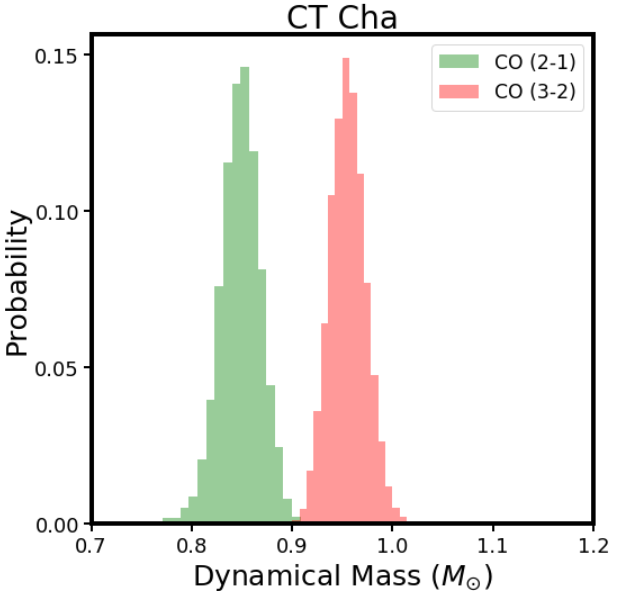} }%
    \qquad
    {{\includegraphics[scale=0.5]{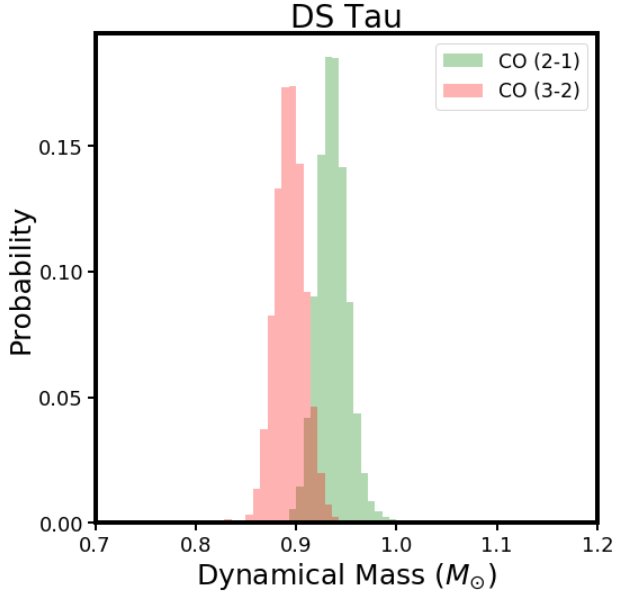} }}%
    \caption{Posterior probability distributions of the central mass from our disk modeling. For CT Cha, the mass derived with CO(3-2) is $0.11 \pm 0.03 ~M_\odot$ (3.7 $\sigma$) higher than of CO(2-1). For DS Tau, however, the mass derived with CO(3-2) is $0.05 \pm 0.02 ~M_\odot$ (2.5 $\sigma$) lower.}
    \label{fig:example}%
\end{figure}

To test whether CO(3-2) also yields a higher mass for a different star, we then repeated the process for DS Tau. We use archival data from ALMA project 2011.0.00150.S (PI: Akeson); details of the observations can be found in \citet{akeson2014}. The velocity resolution is 0.85 km s$^{-1}$ for CO(3-2)  and 1.26 km s$^{-1}$ for CO(2-1). We find that the masses derived from the two transitions are somewhat more consistent with each other. We derive a higher mass of $0.94 \pm 0.02~M_\odot$ from CO(2-1) compared to $0.89 \pm 0.01~M_\odot$ from CO(3-2), and hence the masses differ by about $0.05 \pm 0.02 ~M_\odot$ (2.5 $\sigma$). The right panel of Figure 1 displays the mass distribution of DS Tau. Our CO(3-2) result is $0.06 \pm 0.02~M_\odot$ higher than that found by \citet{lodato2019}, who only used that transition.

\section{Discussion}
We found that different transitions yield different central masses and that there is no obvious pattern between the CO line and the sense of the mass measurement, at least for the two targets we studied in this work. It is clear that at this level of precision, the resulting mass uncertainties are underestimated.  Systematic errors in the modeling therefore appear to dominate over measurement uncertainties. The difference in the derived masses of the two transitions could be due to the assumptions in our radiative transfer models, for example, a power law for the radial gas temperature, and a vertically isothermal structure (\citealt{sheehan2019}). 

Our mass measurement for CT Cha CO(2-1) is higher than \citet{sheehan2019}, despite using the same dataset. They used an older version of \texttt{pdspy}, which assumed that the radial velocity does not change as a function of height above the midplane. As CO is emitted from the disk atmosphere, the radial velocity probed by CO emission should be different from that in the midplane. The radial component of the gravitational force from the star becomes smaller at the surface of a flared disk. So, in order to have the same velocity at the same disk radius, the inferred stellar mass increases to compensate for this, which is reflected in the current version of \texttt{pdspy}. This implies that the uniqueness and systematics of the mass solutions are closely related to how one models the disk.

Our initial results suggest there are systematics that are not well explored, and in the future we plan to probe these effects by measuring the masses for more stars using disk models with detailed temperature structure. We will also characterize the effects of spatial/spectral resolutions and signal-to-noise ratios. Understanding these systematic errors is increasingly relevant in this new era of precise dynamical mass measurements with ALMA and \textit{Gaia}.

We acknowledge the support of the John W. Cox Endowment for advanced studies in astronomy by The Department of Astronomy at The University of Texas at Austin. This paper makes use of the following ALMA data: ADS/JAO.ALMA \#2016.1.01018.S., \#2015.1.00773.S., \#2011.0.00150.S. The authors also acknowledge the Texas Advanced Computing Center (TACC) at The University of Texas at Austin for providing HPC resources that have contributed to the research results reported within this paper. 



\end{document}